\documentclass[oneside,reqno,12pt]{article}
\usepackage{amsmath}

\newif\iffmf
   \fmftrue                                         %
\iffmf
 \usepackage{feynmp}
 \newcommand\fmfbox[2]{%
  \parbox{#1}{%
   \begin{center}
    #2
   \end{center}
  }
 }
\else
 \newcommand\fmfbox[2]{%
  \parbox{#1}{%
   \begin{center}
    (Diagram)
   \end{center}
  }
 }
 \newenvironment{fmffile}[1]{\relax}{\relax}
 \newcommand\fmfcmd[1]{\relax}
\fi

\title{
 Spin-Spin
 Interaction \\
 In Matrix Theory}
\author{
 Miguel Barrio
  \thanks{\texttt{mbarrio@midway.uchicago.edu}}
  \numbermark{1}
 \and
 Robert Helling
  \thanks{\texttt{helling@x4u2.desy.de}}
  \numbermark{2}
 \and
 Gavin Polhemus
  \thanks{\texttt{g-polhemus@uchicago.edu}}
  \numbermark{1}
 }
\date{
  \numbermark{1}
 \textit{
  Department of Physics and Enrico Fermi Institute\\
  The University of Chicago\\
  5640 Ellis Avenue, Chicago, Illinois 60637
 }
 \\
 \bigskip
  \numbermark{2}
 \textit{
  Max Planck Institut f\"ur Gravitationsphysik\\
  Albert-Einstein-Institut\\
  Schlaatzweg 1\\
  14473 Potsdam\\
  Germany
 }
}

\newcommand{\EFInumber}{EFI-98-03}
\newcommand{\archivenumber}{hep-th/9801189}

\newcommand{\numbermark}[1]
 {\renewcommand\thefootnote{\arabic{footnote}}\footnotemark[#1]}

\numberwithin{equation}{section}

\newcommand{\ie}{i.e.\ }

\newcommand{\mean}[1]{\left\langle #1 \right\rangle}

\newcommand{\commute}[2]{\left[ #1,#2 \right]}
\newcommand{\anticommute}[2]{\left\{ #1,#2 \right\}}

\newcommand{\infinity}{\infty}

\newcommand{\textfrac}[2]{{\textstyle\frac{#1}{#2}}}
\newcommand{\half}{\textfrac{1}{2}}

\newcommand{\Tr}{\mathop{\rm Tr}\nolimits}

\newcommand\dt{\partial_{t}}
\newcommand\Dt{D_{t}}
\newcommand\Lagrangian{\mathcal{L}}
\newcommand\action{\mathcal{S}}

\newcommand\cc[1]{\bar{#1}}

\renewcommand\bar[1]{\overline{#1}}

\newcommand\eff{_{\text{eff}}}
\newcommand\MX{\mathbf{X}}
\newcommand\MA{\mathbf{A}}

\newcommand\MO{\boldsymbol{\O}}

\newcommand\cl{_{\text{cl}}}
\newcommand\Xcl{\mathbf{X}\cl}

\newcommand\q{_{\text{q}}}
\newcommand\gs{g_{\text{s}}}

\newcommand\ls{l_{\text{s}}}

\newcommand\rs{r\negmedspace\negmedspace/}
\newcommand\ds{\partial{\mskip -9.5mu}/\,}
\newcommand\drop{^{\vphantom{i}}}
\newcommand\Gt{\widetilde\G}

\newcommand{\greekdef}[3]
 {\def#1{\relax\ifmmode#2\else#3\fi}}
\def\a{\alpha}
\def\b{\beta}
\newcommand\g{\gamma}
\def\d{\delta}
\newcommand\e{\epsilon}

\greekdef\o{\theta}{\char"1C}
\greekdef\i{\iota}{\char"10}

\greekdef\l{\lambda}{\char32l}

\newcommand\p{\pi}

\newcommand\y{\psi}
\newcommand\w{\omega}

\newcommand\G{\Gamma}

\greekdef\O{\Theta}{\char"1F}
\greekdef\L{\Lambda}
 {\leavevmode\setbox0\hbox{L}\hbox to\wd0{\hss\char32L}}

\greekdef\P{\Pi}{\mathhexbox27B}
\greekdef\S{\Sigma}{\mathhexbox278}

\begin{document}
\setlength{\unitlength}{1mm}
\begin{fmffile}{matrixpics7}

\fmfcmd{%
  style_def robertsdings expr p =
    cdraw (wiggly p);
    shrink (1);
      cfill (arrow p);
    endshrink;
  enddef;}

\begin{titlepage}

\maketitle
\thispagestyle{empty}

\begin{table}[t!]
\rightline{\archivenumber}
\rightline{AEI-056}
\rightline{\EFInumber}
\end{table}

\begin{abstract}
We calculate the spin dependent static force between two D0-branes
in Matrix theory.  Supersymmetry relates velocity dependent potentials
to spin dependent potentials.  The well known $v^4\!/r^7$ term is
related to a $\o^8\!/r^{11}$ term, where $\o$ is the
relative spin of the D0-branes.  We calculate this term, confirming that
it is the lowest order contribution to the static potential, and find
its structure consistient with supergravity.
\end{abstract}

\end{titlepage}

\section{Introduction}
\setcounter{footnote}{0}

Matrix theory is conjectured to be M-theory in
the large $N$ limit \cite{Banks:1997vh}.
Therefore, it should contain eleven dimensional supergravity at low energies.
Even prior to this conjecture, it was argued that the leading order
scattering amplitude of two gravitons can be calculated using the ten
dimensional
gauge theory of the string zero modes dimensionally reduced to the
world lines of D0-particles, \ie Matrix theory
\cite{Polchinski:1996fm}.  The spin independent interaction of
D0-branes,proportional to $v^4\!/r^7$, has been calculated and was shown
to be in agreement with eleven dimensional supergravity up to the two loop
level \cite{Douglas:1997yp,Becker:1997wh,Becker:1997xw}.

There should be other terms related to the bosonic $v^4\!/r^7$ term by
supersymmetry \cite{Morales:1997hk,Harvey:1997ic}.  Indeed, the power
counting of the
Matrix theory allows one to trade $\o^2\!/r$ for each power of $v$, where
$\o$ is the relative spin degree of freedom.
The first of these terms, $v^3\o^2\!/r^8$, is a spin-orbit interaction and was
calculated in \cite{Kraus:1997st}, where agreement with supergravity was found.
It has been argued, by comparing massless open string loops to
massless closed string exchange, that all of these terms will agree
\cite{Morales:1997hk}.
In this paper we calculate the static force between D0-branes,
expected to go like $\o^8\!/r^{11}$.

Recently, discrepancies between na\"\i ve matrix predictions and
supergravity were found for three graviton scattering
\cite{Dine:1997sz}, for Matrix theory on ALE spaces
\cite{Douglas:1997uy} and for ${\cal R}^4$ terms in supergravity that
arise from one loop divergences \cite{Keski-Vakkuri:1997wr}.  These
discrepancies are argued to be due to the fact
that these are finite $N$ calculations and that these should be
compared to the compactification of eleven dimensional supergravity on
a light-like circle \cite{Banks:1997mn,Susskind:1997cw,Seiberg:1997ad}.
We have little to add to this debate.
The exact numerical agreement that was miraculously found for the two
graviton amplitude is expected to persist in our calculation since it is
simply a supersymmetric extension of that earlier calculation.
It is our hope that the techniques presented in this paper will
eventually prove useful for calculations involving more than two particles.
These results could help clarify the relation between eleven
dimensional supergravity and Matrix theory.

Calculating the effective potential is equivalent to calculating the
effective action for a static configuration.
There are two techniques for calculating the effective action.  One is
to treat the classical field (defined to be the expectation value of the
quantum field) as a background field \cite{Peskin,Becker:1997wh}.
Quantum corrections come from one-particle-irreducible vacuum diagrams.
The other is to treat the classical field as a perturbation,
expanding in powers of the field \cite{Cheng&Li}.  In this method the quantum
corrections to the $n$th coefficient in the expansion are
one-particle-irreducible diagrams with $n$ external lines.
In this paper we use both methods.  The classical bosonic
fields, representing the positions of the particles, are treated as a
background while the fermionic fields, representing the particles'
spin, are treated as perturbations.
This is fairly natural since the fermionic degrees of freedom are
anticommuting and therefore their power series expansion terminates
after a finite number of terms (at least for finite $N$).

Following this philosophy we find the Feynman rules for the component
fields of $N =2$ Matrix theory.
We show that all terms with $4k+2$ background fermions vanish.
This is related by supersymmetry to the vanishing of odd powers of velocity
in the spin independent case.
Then, we work out the $\o^4\!/r^5$ potential.
This vanishes in agreement to the vanishing of the corresponding bosonic
term $v^2\!/r^3$.
Finally, we find the first nonvanishing static
contribution, which is proportional to $\o^8\!/r^{11}$.

\section{The Bosonic Background}

We start from the Matrix theory action
\begin{equation}
 \action_{\text{Matrix}}
 =
 \frac{1}{\gs\ls}
 \int \! dt \Tr
 \left(
  - (\Dt \MX^i)^2
  + \frac{1}{2\ls^4}\commute{\MX^i}{\MX^j}^2
  - i \MO \Dt\MO
  + \frac{i}{\ls^2} \MO \g^i \commute{\MX^i}{\MO}
 \right),
 \label{eq-Matrix}
\end{equation}
where $\Dt \MX^i  = \dt \MX^i + \commute{\MA}{\MX^i}$.

Following the strategy outlined in the introduction, we split the
bosonic fields into a classical part (obeying the classical equations of
motion) and quantum fluctuations.
The gauge invariance of the background field can be used to set $\MA\cl=0$.
The gauge fixing condition for $\MA\q$ comes from requiring
the background covariant divergence of the gauge fields to vanish.
\begin{equation*}
 \ls^2\dt \MA\q - \frac{1}{\ls^{2}}\commute{\MX\cl^i}{\MX\q^i}
 = 0
\end{equation*}
This leads to a gauge fixing term.
\begin{equation}
 \Lagrangian_{\text{gf}} =
 \frac{1}{\gs\ls} \Tr
 \left(
  \ls^2\dt \MA\q  - \frac{1}{\ls^2}\commute{\Xcl^i}{\MX\q^i}
 \right)^2
\end{equation}
The Lagrangian for the Faddeev-Popov ghosts can be found in
\cite{Becker:1997wh}.
However, it is not needed for the problem under consideration since the ghosts
do not couple to the fermions at 1-loop order.

We separate the Lagrangian into classical and quantum parts.
\begin{equation}
 \Lagrangian
 =
 \frac{1}{\gs}
 (
  \Lagrangian\cl
  + \Lagrangian_X
  + \Lagrangian_A
  + \Lagrangian_\O
  + \Lagrangian_{\text{ghost}}
 )
\end{equation}
We require the background fields satisfy the classical field
equations, so all terms containing a single quantum field vanish.
We also use the identity
\begin{equation*}
 \Tr \commute{\MX\cl^i}{\MX\q^j}\commute{\MX\q^i}{\MX\cl^j}
 + \Tr \commute{\MX\cl^i}{\MX\q^i}^2
 =
 \Tr \commute{\MX\cl^i}{\MX\cl^j}\commute{\MX\q^i}{\MX\q^j},
\end{equation*}
which can be derived from  $\Tr A\commute BC = \Tr\commute AB C$ and
the Jacobi identity.
\begin{equation}
 \begin{split}
  \Lagrangian_X
  &=
  \Tr
  \biggl(
   - (\dt \MX\q^i)^2
   + \commute{\MX\cl^i}{\MX\q^j}^2
   + 2\commute{\MX\cl^i}{\MX\cl^j}\commute{\MX\q^i}{\MX\q^j}
   \\ & \qquad \qquad \qquad \qquad \qquad \qquad
   + 2\commute{\MX\cl^i}{\MX\q^j}\commute{\MX\q^i}{\MX\q^j}
   + \frac{1}{2}\commute{\MX\q^i}{\MX\q^j}^2
  \biggr)
  \\
  \Lagrangian_A
  & =
  \Tr
  \biggl(
   (\dt \MA)^2
   - 4i \dt \MX\cl^i \commute{\MA}{\MX\q^i}
   - \commute{\MA}{\MX\cl^i}^2
   \\ & \qquad \qquad \qquad
   - 2i \dt \MX\q^i \commute{\MA}{\MX\q^i}
   - 2\commute{\MA}{\MX\cl^i}\commute{\MA}{\MX\q^i}
   - \commute{\MA}{\MX\q^i}^2
  \biggr)
  \\
  \Lagrangian_\O
  &=
  \Tr
  \left(
   - i \MO \dt\MO
   - i \MO \commute{\MA}{\MO}
   + i \MO \g^i \commute{\MX\cl^i}{\MO}
   + i \MO \g^i \commute{\MX\q^i}{\MO}
  \right)
 \end{split}
\end{equation}

\subsection{Component Fields for $N=2$}
We are interested in the static force between two D0-branes in the
center of mass system. Therefore we consider only a static background
of fields in the Cartan subalgebra.

\begin{align*}
 \Xcl^i &= \frac{i}{2}
  \begin{bmatrix}
   r^i &  0   \\
   0   & -r^i
  \end{bmatrix}
 &
 \MX\q^i &= \frac{i}{2}
  \begin{bmatrix}
   X^i & \sqrt{2}\,\cc Y^i \\
   \sqrt{2}\,Y^i & -X^i
  \end{bmatrix}
 \\
 \MA &= \frac{i}{2}
  \begin{bmatrix}
   A & \sqrt{2}\,\cc B \, \\
   \sqrt{2}\,B & -A
  \end{bmatrix}
 &
 \MO &= \frac{i}{2}
  \begin{bmatrix}
   \o & \sqrt{2}\,\bar \y \, \\
   \sqrt{2}\,\y & -\o
  \end{bmatrix}
\end{align*}
The bar represents complex conjugation.  All of the fields on the
diagonals are real, while the off diagonal fields are complex.

The Lagrangian can be written in terms of these real and complex
fields.  Only interactions involving fermions will be necessary for
this calculation, so purely bosonic interactions are left out of
the Lagrangians.
\begin{gather}
 \Lagrangian_X =
 - \half X^i \dt^2 X^i - \cc{Y}^i (\dt^2 + r^2) Y^i
 + \text{(interactions)}
 \notag \\
 \Lagrangian_A =
 \half A \dt^2 A + \cc{B} (\dt^2 + r^2) B
 + \text{(interactions)}
 \\
 \Lagrangian_\O =
 \half \o i \dt \o + \bar\y (i\dt - \rs) \y
 + Y^i \bar\y \g^i \o + \cc Y^i \o \g^i \y
 - B \bar\y \o - \cc B \o \y
 - X^i \bar\y \g^i \y + A \bar\y \y
 \notag
\end{gather}
We will integrate out the off diagonal fields to get an effective action
for the diagonal fields.  The background has given a mass to
the off diagonal fields, which will prevent infrared divergences.

\subsection{Feynman rules}
The Feynman rules can easily be read off of the Lagrangian.
The off diagonal fields will be integrated out,
so only their propagators are needed.
\begin{gather*}
 \mean{\cc Y^i Y^j}
 =
 \fmfbox{25mm}{
  \begin{fmfchar*}(15,10)
   \fmfleft{l}
   \fmfright{r}
   \fmf{robertsdings}{l,r}
   \fmflabel{$i$}{r}
   \fmflabel{$j$}{l}
  \end{fmfchar*}
 }
 = \frac{i\d^{ij}}{\w^2 - r^2 +i\e}
 \qquad
 \mean{\cc B B}
 =
\fmfbox{25mm}{
 \begin{fmfchar*}(25,10)
  \fmfleft{l}
  \fmfright{r}
  \fmf{scalar}{l,r}
 \end{fmfchar*}
}
 = \frac{-i}{\w^2 - r^2 +i\e}
 \\
 \mean{\bar \y \y}
 =
 \fmfbox{25mm}{
  \begin{fmfchar*}(15,10)
   \fmfleft{l}
   \fmfright{r}
   \fmf{fermion}{l,r}
   \fmflabel{$\a$}{r}
   \fmflabel{$\b$}{l}
  \end{fmfchar*}
 }
 = i\frac{(\w + \rs)_{\a\b}}{\w^2 - r^2 +i\e}
\end{gather*}

The spin of the D0-branes enters the calculation through the
interaction of the above massive fields with the $\o$ field.  The
vertices (with the external $\o$ attached, as explained below) are
\begin{align*}
 \fmfbox{35mm}{
  \begin{fmfchar*}(35,10)
   \fmfbottom{lo,ro}
   \fmftop{t}
   \fmfv{decor.shape=cross,decor.filled=full,decor.size=5thick}{t}
   \fmfdot{m}
   \fmf{phantom,tension=3}{lo,l}
   \fmf{phantom,tension=3}{r,ro}
   \fmf{robertsdings}{l,m}
   \fmf{fermion}{m,r}
   \fmf{plain,tension=.5}{t,m}
   \fmflabel{$\a$}{r}
   \fmflabel{$i$}{l}
  \end{fmfchar*}
 }
 &= i \g^i_{\a\b} \o_\b\drop
 &
 \fmfbox{35mm}{
  \begin{fmfchar*}(35,10)
   \fmfbottom{lo,ro}
   \fmftop{t}
   \fmfv{decor.shape=cross,decor.filled=full,decor.size=5thick}{t}
   \fmfdot{m}
   \fmf{phantom,tension=3}{lo,l}
   \fmf{phantom,tension=3}{r,ro}
   \fmf{fermion}{l,m}
   \fmf{robertsdings}{m,r}
   \fmf{plain,tension=.5}{t,m}
   \fmflabel{$i$}{r}
   \fmflabel{$\a$}{l}
  \end{fmfchar*}
 }
 &= i \o_\b\drop \g^i_{\b\a}
 \\
 \fmfbox{35mm}{
  \begin{fmfchar*}(35,10)
   \fmfbottom{l,ro}
   \fmftop{t}
   \fmfv{decor.shape=cross,decor.filled=full,decor.size=5thick}{t}
   \fmfdot{m}
   \fmf{phantom,tension=3}{r,ro}
   \fmf{scalar,tension=.75}{l,m}
   \fmf{fermion}{m,r}
   \fmf{plain,tension=.5}{t,m}
   \fmflabel{$\a$}{r}
  \end{fmfchar*}
 }
 &= -i \o_\a &
 \fmfbox{35mm}{
  \begin{fmfchar*}(35,10)
   \fmfbottom{lo,r}
   \fmftop{t}
   \fmfv{decor.shape=cross,decor.filled=full,decor.size=5thick}{t}
   \fmfdot{m}
   \fmf{phantom,tension=3}{lo,l}
   \fmf{fermion}{l,m}
   \fmf{scalar,tension=.75}{m,r}
   \fmf{plain,tension=.5}{t,m}
   \fmflabel{$\a$}{l}
  \end{fmfchar*}
 }
 &= -i \o_\a.
\end{align*}

\section{The Expansion in $\o$}

As described in the introduction, we expand the effective action in
powers of $\o$.
\begin{equation}
 \G(r,\o)
 =
 \sum_n \int dt_1 \dotsm dt_n\,
 \G^{(n)}_{\a_1,\dots,\a_n}\!(r; t_1, \dots, t_n) \,
 \o_{\a_1}(t_1) \dotsm \o_{\a_n}(t_n)
 \label{expansion}
\end{equation}
It turns out that $i\G^{(n)}_{\a_1,\dots,\a_n}(r; t_1, \dots, t_n)$ is
the sum of all one-particle-irreducible diagrams with $n$ external
fermion lines.  In frequency space, the effective potential is
just minus the effective action evaluated at zero momentum:
\begin{equation}
 \begin{split}
  V\eff(r,\o)
  &=
  - \sum_n
  \Gt^{(n)}_{\a_1,\dots,\a_n}\!(r;0,\dots,0)\,
  \o_{\a_1}\!\dotsm\o_{\a_n}
  \\
  &=
  i \sum_n \text{(diagrams)}
  \label{Veff&Diagrams}
 \end{split}
\end{equation}
where the diagrams are the one-particle-irreducible diagrams with
$n$, zero-frequency external lines with $\o$'s attached.

\subsection{Terms Proportional to $\o^{4k+2}$}

The $\o^2$ term is easy, since $\o \o = \o\g^i\o =0 $.  First, the
diagram with the gauge field.
\begin{equation}
 \fmfbox{32mm}{
  \begin{fmfchar*}(32,24)
   \fmfleft{tl}
   \fmfright{tr}
   \fmfv{decor.shape=cross,decor.filled=full,decor.size=5thick}{tl,tr}
   \fmfdot{l,r}
   \fmf{plain,tension=10}{tl,l}
   \fmf{plain,tension=10}{tr,r}
   \fmf{scalar,left,tension=1}{l,r}
   \fmf{fermion,left,tension=1}{r,l}
  \end{fmfchar*}
 }
 \; =
 \int_{-\infty}^\infty \frac{d\w}{2\pi}
 \frac{-\o (\w + \rs) \o}{(\w^2 - r^2 + i\epsilon)^2}
 = 0
\end{equation}
In fact any diagram that has a fermion between two gauge
fields vanishes.
Next, the one with the $Y$ fields.
\begin{equation}
 \fmfbox{32mm}{
  \begin{fmfchar*}(32,24)
   \fmfleft{tl}
   \fmfright{tr}
   \fmfv{decor.shape=cross,decor.filled=full,decor.size=5thick}{tl,tr}
   \fmfdot{l,r}
   \fmf{plain,tension=10}{tl,l}
   \fmf{plain,tension=10}{tr,r}
   \fmf{robertsdings,left,tension=1}{l,r}
   \fmf{fermion,left,tension=1}{r,l}
  \end{fmfchar*}
 }
 \; =
 \int_{-\infty}^\infty \frac{d\w}{2\pi}
 \frac{\o \g^i (\w + \rs) \g^i \o}{(\w^2 - r^2 + i\epsilon)^2}
 = 0
\end{equation}
Both of the above diagrams are traversed by only one fermion.
We will call any series of fermion traversals connected by vectors ($Y$
fields) a \emph{chain}.  A chain is ended by scalars ($B$ fields), or
it connects back on its self
(when the loop contains no scalars).  As an example, there are four $\o^6$
diagrams:

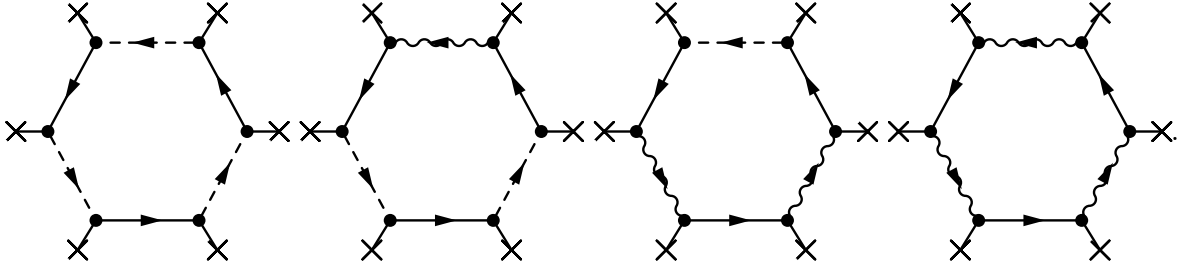
\begin{equation*}
 \fmfbox{35mm}{
 \begin{fmfchar*}(35,35)
  \fmfpen{thin}
  \fmfset{arrow_len}{3mm}
  \fmfsurroundn{t}{6}
  \begin{fmffor}{n}{1}{1}{6}
   \fmfv{decor.shape=cross,decor.filled=full,decor.size=5thick}{t[n]}
   \fmfdot{i[n]}
   \fmf{plain,tension=3}{t[n],i[n]}
  \end{fmffor}
  \fmf{fermion}{i1,i2}
  \fmf{fermion}{i3,i4}
  \fmf{fermion}{i5,i6}
  \fmf{scalar}{i2,i3}
  \fmf{scalar}{i4,i5}
  \fmf{scalar}{i6,i1}
 \end{fmfchar*}
 }
 \quad
 \fmfbox{35mm}{
 \begin{fmfchar*}(35,35)
  \fmfpen{thin}
  \fmfset{arrow_len}{3mm}
  \fmfsurroundn{t}{6}
  \begin{fmffor}{n}{1}{1}{6}
   \fmfv{decor.shape=cross,decor.filled=full,decor.size=5thick}{t[n]}
   \fmfdot{i[n]}
   \fmf{plain,tension=3}{t[n],i[n]}
  \end{fmffor}
  \fmf{fermion}{i1,i2}
  \fmf{fermion}{i3,i4}
  \fmf{fermion}{i5,i6}
  \fmf{robertsdings}{i2,i3}
  \fmf{scalar}{i4,i5}
  \fmf{scalar}{i6,i1}
 \end{fmfchar*}
 }
 \quad
 \fmfbox{35mm}{
  \begin{fmfchar*}(35,35)
   \fmfpen{thin}
   \fmfset{arrow_len}{3mm}
   \fmfsurroundn{t}{6}
   \begin{fmffor}{n}{1}{1}{6}
    \fmfv{decor.shape=cross,decor.filled=full,decor.size=5thick}{t[n]}
    \fmfdot{i[n]}
    \fmf{plain,tension=3}{t[n],i[n]}
   \end{fmffor}
   \fmf{fermion}{i1,i2}
   \fmf{fermion}{i3,i4}
   \fmf{fermion}{i5,i6}
   \fmf{scalar}{i2,i3}
   \fmf{robertsdings}{i4,i5}
   \fmf{robertsdings}{i6,i1}
  \end{fmfchar*}
 }
 \quad
 \fmfbox{35mm}{
  \begin{fmfchar*}(35,35)
   \fmfpen{thin}
   \fmfset{arrow_len}{3mm}
   \fmfsurroundn{t}{6}
   \begin{fmffor}{n}{1}{1}{6}
    \fmfv{decor.shape=cross,decor.filled=full,decor.size=5thick}{t[n]}
    \fmfdot{i[n]}
    \fmf{plain,tension=3}{t[n],i[n]}
   \end{fmffor}
   \fmf{fermion}{i1,i2}
   \fmf{fermion}{i3,i4}
   \fmf{fermion}{i5,i6}
   \fmf{robertsdings}{i2,i3}
   \fmf{robertsdings}{i4,i5}
   \fmf{robertsdings}{i6,i1}
  \end{fmfchar*}
 }\;.
\end{equation*}
The first contains three chains with only a single fermion traversing
each (one link chains).  The second contains a one link and a two link chain.
The last two contain three link chains.

Here we prove that any diagram containing a chain with an
odd number of links is zero.  First, consider a closed chain with $n$
links.  Before doing the integral over $\w$, it will contain a factor
\begin{equation*}
 \o \g^{i_1} (\w + \rs) \g^{i_2} \o \,
 \o \g^{i_2} (\w + \rs) \g^{i_3} \o \,
 \dotsb
 \o \g^{i_{n-1}} (\w + \rs) \g^{i_{n}} \o \,
 \o \g^{i_{n}} (\w + \rs) \g^{i_1} \o.
\end{equation*}
Each fermion bilinear is antisymmetric in its vector indices.  Since there
are an odd number of these factors, swapping the indices on all of
them produces an overall minus sign.
\begin{equation*}
 -\o \g^{i_2} (\w + \rs) \g^{i_1} \o \,
 \o \g^{i_3} (\w + \rs) \g^{i_2} \o \,
 \dotsb
 \o \g^{i_{n}} (\w + \rs) \g^{i_{n-1}} \o \,
 \o \g^{i_1} (\w + \rs) \g^{i_{n}} \o
\end{equation*}
The factors can then be reordered,
\begin{equation*}
 -\o \g^{i_1} (\w + \rs) \g^{i_{n}} \o \,
 \o \g^{i_{n}} (\w + \rs) \g^{i_{n-1}} \o \,
 \dotsb
 \o \g^{i_3} (\w + \rs) \g^{i_2} \o \,
 \o \g^{i_2} (\w + \rs) \g^{i_1} \o,
\end{equation*}
to reproduce the original expression, but with a minus sign (and the
indices renamed).
This implies that the term is zero.  Chains ended by scalars
contribute a factor
\begin{equation*}
 \o \rs \g^{i_1} \o \,
 \o \g^{i_1} (\w + \rs) \g^{i_2} \o \,
 \dotsb
 \o \g^{i_{n-2}} (\w + \rs) \g^{i_{n-1}} \o \,
 \o \g^{i_{n-1}} \rs \o
\end{equation*}
which vanishes for similar reasons.
Therefore, any diagram containing a chain with an odd number of links is zero.
All diagrams that are order $2 \pmod{4}$ in $\o$ have an odd number of
fermions
traversing them, so they must contain a chain with an odd number of links.
For a diagram to give a nonvanishing contribution, the number of
external $\o$ lines must be a multiple of four.
The diagrams with $\theta^{4k+2}$ are related by supersymmetry to
bosonic diagrams with $v^{2k+1}$, which are required to vanish by
time-reversal symmetry.

\subsection{The Vanishing $\o^4$ Term}

The $\o^4$ term is also zero, but demonstrating this is more difficult.
There is one diagram that vanishes because it contains odd chains.
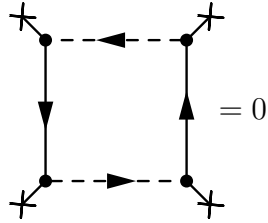
\begin{equation*}
 \fmfbox{25mm}{
  \begin{fmfchar*}(25,25)
   \fmfstraight
   \fmfleft{t1,t2}
   \fmfright{t4,t3}
   \fmfdot{lo,lu,ro,ru}
   \fmfv{
    decor.shape=cross,decor.filled=full,
    decor.size=5thick,decor.angle=40
   }{t1,t2,t3,t4}
   \fmf{plain,tension=6}{t1,lo}
   \fmf{plain,tension=6}{t2,lu}
   \fmf{plain,tension=6}{t3,ru}
   \fmf{plain,tension=6}{t4,ro}
   \fmf{fermion,tension=1}{lu,lo}
   \fmf{fermion,tension=1}{ro,ru}
   \fmf{scalar}{lo,ro}
   \fmf{scalar}{ru,lu}
  \end{fmfchar*}
 }
 = 0
\end{equation*}
The cancellation of the remaining two diagrams requires two identities,
  \eqref{identity41} and \eqref{identity42},
which are derived in the appendix.
\begin{equation}
 \begin{split}
  \smash[b]{
   \fmfbox{25mm}{
    \begin{fmfchar*}(25,25)
     \fmfleft{t1,t2}
     \fmfright{t4,t3}
     \fmfdot{lo,lu,ro,ru}
     \fmfv{
      decor.shape=cross,decor.filled=full,
      decor.size=5thick,decor.angle=40
     }{t1,t2,t3,t4}
     \fmf{plain,tension=6}{t1,lo}
     \fmf{plain,tension=6}{t2,lu}
     \fmf{plain,tension=6}{t3,ru}
     \fmf{plain,tension=6}{t4,ro}
     \fmf{fermion,tension=1}{lu,lo}
     \fmf{fermion,tension=1}{ro,ru}
     \fmf{robertsdings}{lo,ro}
     \fmf{scalar}{ru,lu}
    \end{fmfchar*}
   }
  }
  \quad &=
  -\int_{-\infty}^\infty \frac{d\w}{2\pi}
  \frac{
   \o (\w + \rs) \g^i \o \,
   \o \g^i (\w + \rs) \o
  }{(\w^2-r^2+i\epsilon)^4}
  \\
  &=
  -\int_{-\infty}^\infty \frac{d\w}{2\pi}
  \frac{
   \o \rs \g^i \o \,
   \o \g^i \rs \o
  }{(\w^2-r^2+i\epsilon)^4}
 \end{split}
\end{equation}
\begin{equation}
 \begin{split}
  \smash[b]{
   \fmfbox{25mm}{
    \begin{fmfchar*}(25,25)
     \fmfleft{t1,t2}
     \fmfright{t4,t3}
     \fmfv{
      decor.shape=cross,decor.filled=full,
      decor.size=5thick,decor.angle=40
     }{t1,t2,t3,t4}
     \fmfdot{lo,lu,ro,ru}
     \fmf{plain,tension=6}{t1,lo}
     \fmf{plain,tension=6}{t2,lu}
     \fmf{plain,tension=6}{t3,ru}
     \fmf{plain,tension=6}{t4,ro}
     \fmf{fermion,tension=1}{lu,lo}
     \fmf{fermion,tension=1}{ro,ru}
     \fmf{robertsdings}{lo,ro}
     \fmf{robertsdings}{ru,lu}
    \end{fmfchar*}
   }
  }
  \quad &=
  \frac{1}{2}
  \int_{-\infty}^\infty \frac{d\w}{2\pi}
  \frac{
   \o \g^i (\w + \rs) \g^j \o \,
   \o \g^j (\w + \rs) \g^i \o
  }{(\w^2-r^2+i\epsilon)^4}
  \\
  &=
  \frac{1}{2}
  \int_{-\infty}^\infty \frac{d\w}{2\pi}
  \frac{
   \w^2
   \o \g^i \g^j \o \,
   \o \g^j \g^i \o
   +
   \o \g^i \rs \g^j \o \,
   \o \g^j \rs \g^i \o
  }{(\w^2-r^2+i\epsilon)^4}
  \\
  &=
  \frac{1}{2}
  \int_{-\infty}^\infty \frac{d\w}{2\pi}
  \frac{
   \o \g^i \rs \g^j \o \,
   \o \g^j \rs \g^i \o
  }{(\w^2-r^2+i\epsilon)^4}
  \\
  &=
  \int_{-\infty}^\infty \frac{d\w}{2\pi}
  \frac{
   \o \rs \g^i \o \,
   \o \g^i \rs \o
  }{(\w^2-r^2+i\epsilon)^4}
 \end{split}
\end{equation}
Again, the vanishing of this term is related via supersymmetry to
the vanishing of a purely bosonic term, namely $v^2\!/r^5$.

\subsection{The $\o^8$ Term}

The $\o^8$ term should give the first non-zero
contribution since it corresponds to the well known $v^4\!/r^7$
term.  There are three diagrams that vanish due to odd
chains:
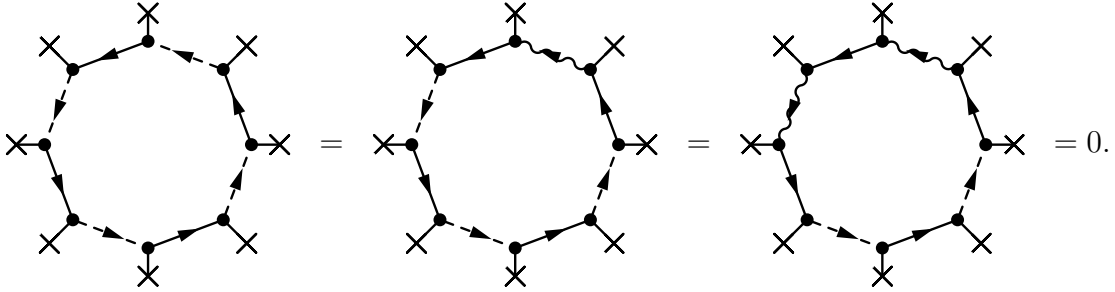
\begin{equation*}
\fmfbox{35mm}{
 \begin{fmfchar*}(35,35)
  \fmfcurved
  \fmfpen{thin}
  \fmfset{arrow_len}{3mm}
  \fmfsurroundn{t}{8}
  \begin{fmffor}{n}{1}{1}{8}
   \fmfv{decor.shape=cross,decor.filled=full,decor.size=5thick}{t[n]}
   \fmfdot{i[n]}
   \fmf{plain,tension=2}{t[n],i[n]}
  \end{fmffor}
  \fmf{fermion}{i1,i2}
  \fmf{fermion}{i3,i4}
  \fmf{fermion}{i5,i6}
  \fmf{fermion}{i7,i8}
  \fmf{scalar}{i2,i3}
  \fmf{scalar}{i4,i5}
  \fmf{scalar}{i6,i7}
  \fmf{scalar}{i8,i1}
 \end{fmfchar*}
}
\quad = \quad
\fmfbox{35mm}{
 \begin{fmfchar*}(35,35)
  \fmfpen{thin}
  \fmfset{arrow_len}{3mm}
  \fmfsurroundn{t}{8}
  \begin{fmffor}{n}{1}{1}{8}
   \fmfv{decor.shape=cross,decor.filled=full,decor.size=5thick}{t[n]}
   \fmfdot{i[n]}
   \fmf{plain,tension=2}{t[n],i[n]}
  \end{fmffor}
  \fmf{fermion}{i1,i2}
  \fmf{fermion}{i3,i4}
  \fmf{fermion}{i5,i6}
  \fmf{fermion}{i7,i8}
  \fmf{robertsdings}{i2,i3}
  \fmf{scalar}{i4,i5}
  \fmf{scalar}{i6,i7}
  \fmf{scalar}{i8,i1}
 \end{fmfchar*}
}
\quad = \quad
\fmfbox{35mm}{
 \begin{fmfchar*}(35,35)
  \fmfpen{thin}
  \fmfset{arrow_len}{3mm}
  \fmfsurroundn{t}{8}
  \begin{fmffor}{n}{1}{1}{8}
   \fmfv{decor.shape=cross,decor.filled=full,decor.size=5thick}{t[n]}
   \fmfdot{i[n]}
   \fmf{plain,tension=2}{t[n],i[n]}
  \end{fmffor}
  \fmf{fermion}{i1,i2}
  \fmf{fermion}{i3,i4}
  \fmf{fermion}{i5,i6}
  \fmf{fermion}{i7,i8}
  \fmf{robertsdings}{i2,i3}
  \fmf{robertsdings}{i4,i5}
  \fmf{scalar}{i6,i7}
  \fmf{scalar}{i8,i1}
 \end{fmfchar*}
}
\quad = 0.
\end{equation*}

To calculate the remaining diagrams we will need the following
integrals.
\begin{align*}
 \int_{-\infinity}^{\infinity}\frac{d\w}{2\p}
 \frac{1}{(\w^2 - r^2 + i\e)^8}
 &\stackrel{\e \rightarrow 0}{=}
 \frac{429i}{4096 r^{15}}
 \\
 \int_{-\infinity}^{\infinity}\frac{d\w}{2\p}
 \frac{\w^2}{(\w^2 - r^2 + i\e)^8}
 &\stackrel{\e \rightarrow 0}{=}
 -\frac{33i}{4096 r^{13}}
 \\
 \int_{-\infinity}^{\infinity}\frac{d\w}{2\p}
 \frac{\w^4}{(\w^2 - r^2 + i\e)^8}
 &\stackrel{\e \rightarrow 0}{=}
 \frac{9i}{4096 r^{11}}
\end{align*}
The first diagram contains two $B$ and two $Y^i$ fields in the
loop.
\begin{equation}
 \begin{split}
  \smash[b]{
   \fmfbox{35mm}{
    \begin{fmfchar*}(35,35)
     \fmfpen{thin}
     \fmfset{arrow_len}{3mm}
     \fmfsurroundn{t}{8}
     \begin{fmffor}{n}{1}{1}{8}
      \fmfv{decor.shape=cross,decor.filled=full,decor.size=5thick}{t[n]}
      \fmfdot{i[n]}
      \fmf{plain,tension=2}{t[n],i[n]}
     \end{fmffor}
     \fmf{fermion}{i1,i2}
     \fmf{fermion}{i3,i4}
     \fmf{fermion}{i5,i6}
     \fmf{fermion}{i7,i8}
     \fmf{robertsdings}{i2,i3}
     \fmf{scalar}{i4,i5}
     \fmf{robertsdings}{i6,i7}
     \fmf{scalar}{i8,i1}
    \end{fmfchar*}
   }
  }
  \; &=
  \frac{1}{2}
  \int_{-\infinity}^{\infinity}\frac{d\w}{2\p}
  \frac{
   \left(
    \o \rs \g^i \o \, \o \g^i \rs \o
   \right)^2
  }
  {(\w^2 - r^2 + i\e)^8}
  \\
  &=
  \frac{429i}{8192 r^{15}}
  \left(
   \o \rs \g^i \o \, \o \g^i \rs \o
  \right)^2
 \end{split}
\end{equation}

To simplify the two final diagrams we employ
  identities \eqref{identity81}, \eqref{identity82}, \eqref{identity83}
  and \eqref{identity84} from the appendix.
With these we obtain
\begin{equation}
 \begin{split}
  \smash[b]{
   \fmfbox{35mm}{
    \begin{fmfchar*}(35,35)
     \fmfpen{thin}
     \fmfset{arrow_len}{3mm}
     \fmfsurroundn{t}{8}
     \begin{fmffor}{n}{1}{1}{8}
      \fmfv{decor.shape=cross,decor.filled=full,decor.size=5thick}{t[n]}
      \fmfdot{i[n]}
      \fmf{plain,tension=2}{t[n],i[n]}
     \end{fmffor}
     \fmf{fermion}{i1,i2}
     \fmf{fermion}{i3,i4}
     \fmf{fermion}{i5,i6}
     \fmf{fermion}{i7,i8}
     \fmf{robertsdings}{i2,i3}
     \fmf{robertsdings}{i4,i5}
     \fmf{robertsdings}{i6,i7}
     \fmf{scalar}{i8,i1}
    \end{fmfchar*}
   }
  }
  \; &=
  -\int_{-\infinity}^{\infinity}\frac{d\w}{2\p}
  \frac{
   \o \rs \g^i \o \,
   \o \g^i (\w + \rs) \g^j \o \,
   \o \g^j (\w + \rs) \g^k \o \,
   \o \g^k \rs \o
  }
  {(\w^2 - r^2 + i\e)^8}
  \\
  &=
  \frac{33i}{4096r^{13}}
  \o \rs \g^i \o \, \o \g^i \g^j \o \, \o \g^j \g^k \o \, \o \g^k \rs \o
  \\ & \qquad
  -\frac{429i}{4096r^{15}}
  \o \rs \g^i \o \, \o \g^i \rs \g^j \o \,
  \o \g^j \rs \g^k \o \, \o \g^k \rs \o
  \\
  &=
  -\frac{99i}{1024r^{13}}
  \o \rs \g^i \o \, \o \g^i \g^j \o \, \o \g^j \g^k \o \, \o \g^k \rs \o
  \\ & \qquad
  +\frac{429i}{1024r^{15}}
  \left( \o \rs \g^i \o \, \o \g^i \rs \o \right)^2,
 \end{split}
\end{equation}
and finally,
\begin{equation}
 \begin{split}
  \smash[b]{
   \fmfbox{35mm}{
    \begin{fmfchar*}(35,35)
     \fmfpen{thin}
     \fmfset{arrow_len}{3mm}
     \fmfsurroundn{t}{8}
     \begin{fmffor}{n}{1}{1}{8}
      \fmfv{decor.shape=cross,decor.filled=full,decor.size=5thick}{t[n]}
      \fmfdot{i[n]}
      \fmf{plain,tension=2}{t[n],i[n]}
     \end{fmffor}
     \fmf{fermion}{i1,i2}
     \fmf{fermion}{i3,i4}
     \fmf{fermion}{i5,i6}
     \fmf{fermion}{i7,i8}
     \fmf{robertsdings}{i2,i3}
     \fmf{robertsdings}{i4,i5}
     \fmf{robertsdings}{i6,i7}
     \fmf{robertsdings}{i8,i1}
    \end{fmfchar*}
   }
  }
  \; &=
  \frac{1}{4}
  {\int_{-\infinity}^{\infinity}}\frac{d\w}{2\p}
  \frac{
   \o \g^i (\w + \rs) \g^j \o \,
   \o \g^j (\w + \rs) \g^k \o \,
   \o \g^k (\w + \rs) \g^l \o \,
   \o \g^l (\w + \rs) \g^i \o
  }
  {
   (\w^2 - r^2 + i\e)^8
  }
  \mskip -50mu
  \\
  &=
  \frac{9i}{16384r^{11}}
  \o \g^i \g^j \o \, \o \g^j \g^k \o \, \o \g^k \g^l \o \, \o \g^l \g^i \o
  \\ & \qquad
  -
  \frac{33i}{4096r^{13}}
  \o \g^i \g^j \o \, \o \g^j \rs \g^k \o \,
  \o \g^k \rs \g^l \o \, \o \g^l \g^i \o
  \\ & \qquad\qquad
  -
  \frac{33i}{8192r^{13}}
  \o \g^i \rs \g^j \o \, \o \g^j \g^k \o \,
  \o \g^k \rs \g^l \o \, \o \g^l \g^i \o
  \\ & \qquad\qquad\qquad
  +
  \frac{429i}{16384r^{15}}
  \o \g^i \rs \g^j \o \, \o \g^j \rs \g^k \o \,
  \o \g^k \rs \g^l \o \, \o \g^l \rs \g^i \o
  \\
  &=
  \frac{15i}{1024r^{11}}
  \o \g^i \g^j \o \, \o \g^j \g^k \o \, \o \g^k \g^l \o \, \o \g^l \g^i \o
  \\ & \qquad
  -
  \frac{231i}{1024r^{13}}
  \o \rs \g^i \o \, \o \g^i \g^j \o \,
  \o \g^j \g^k \o \, \o \g^k \rs \o
  \\ & \qquad\qquad
  +
  \frac{4719i}{8192r^{15}}
  \left( \o \rs \g^i\o \, \o \g^i \rs \o \right)^2
 \end{split}
\end{equation}

\section{Conclusion}

Summing the above diagrams as in \eqref{Veff&Diagrams} we find the effective
potential at order $\o^8\!/r^{11}$:
\begin{equation}
 \begin{split}
 V\eff(r,\o)
  &=
  -\frac{15}{(2r)^{11}}
  \Biggl(
   2
   \o \g^i \g^j \o \, \o \g^j \g^k \o \, \o \g^k \g^l \o \, \o \g^l \g^i \o
   \\ & \qquad \qquad \qquad
   -
   \frac{44}{r^2}
   \o \rs \g^i \o \, \o \g^i \g^j \o \,
   \o \g^j \g^k \o \, \o \g^k \rs \o
   \\ & \qquad \qquad \qquad \qquad
   +
   \frac{143}{r^4}
   \left( \o \rs \g^i\o \, \o \g^i \rs \o \right)^2
  \Biggr)
  \\
  &=
  -\frac{5}{43{,}008}
  \left(
   \o \ds \g^i\o \, \o \g^i \ds \o
  \right)^2
  \frac{1}{r^7}
 \end{split}
\end{equation}
There may be terms higher order in $\o$.  Since $\o$ only has sixteen
components,
the only two terms that could remain would be proportional to
$\o^{12}\!/r^{17}$ and $\o^{16}\!/r^{23}$.  These are not related to
the $v^4\!/r^7$ term but to higher order $v^6\!/r^{11}$ and $v^8\!/r^{15}$
terms respectively.
It is remarkable that the contributions from the diagrams conspire to give
exactly the coefficient that one gets from acting on $1/r^7$ with four
gradients.

In order to make contact with supergravity we take the Fourier transform.
\begin{equation}
 V\eff(q,\o)
 =
 -\frac{\p^4}{252}
 \frac{(q^i J^{ij}J^{jk} q^k)^2}{q^2}
\end{equation}
where we have put in the
the relative angular momentum of the D0-branes,
$J^{ij} = \frac{i}{2} \o\g^i\g^j\o$
\ \cite{Kraus:1997st}.
The $q^2$ in the denominator is characteristic for an exchanged graviton,
and the structure is the same as the (eight dimensional) static supergravity
result in \cite{Harvey:1997ic}.

We would like to thank J. Harvey for many useful discussions and P.
Pouliot for a discussion of supersymmetry's role in relating the various
potentials.
After the completion of this work, a similar question was addressed using
string scattering theory \cite{Morales:1998}.

\appendix

\section{SO(9) Spinor Identities}
Using the Clifford algebra relation
$\anticommute{\g^i}{\g^j}=2\d^{ij}$ one immediately obtains
\begin{gather*}
 \anticommute{\rs}{\g^i} = 2 r^i
 \qquad \qquad
 \rs\rs = r^2
 \\
 \g^i \g^{j_1 j_2 \dots j_n} \g^i
 =
 (-1)^n (9-2n) \g^{j_1 j_2 \dots j_n},
\end{gather*}
where
$\g^{i_1 i_2 \dots i_n} \equiv \g^{[i_1} \g^{i_2} \dotsm \g^{i_n]}$.

We use a representation of the $SO(9)$ Clifford algebra with real,
symmetric Dirac matrices \cite{GSW}. Therefore, antisymmetrized products of two
and three $\g$-matrices are antisymmetric in the spinor indices whereas
products of
0,1,4, and 5 are symmetric. From this it follows
\begin{alignat}{2}
 \o\o &= 0 &
 \o\g^i\o &= 0
 \notag \\
 \o\g^i\g^j\o &= \o\g^{ij}\o &
 \o\g^i\g^j\g^k\o &= \o\g^{ijk}\o
 \\
 \o\g^i\g^j\g^k\g^l\o
 &=
 \d^{ij}\o\g^{kl}\o
 -\d^{ik}\o\g^{jl}\o
 +\d^{il}\o\g^{jk}\o
 +\d^{jk}\o\g^{il}\o
 -\d^{jl}\o\g^{ik}\o
 +\d^{kl}\o\g^{ij}\o
 \hspace{-4cm}
 \notag
\end{alignat}

We also make frequent use of the following Fierz identity, which can be
derived
from the fact that $\g^{ij}$ and $\g^{ijk}$ form a complete basis for
$16 \times 16$ matrices antisymmetric in $\a$ and $\b$.
\begin{equation}
 \o_\a \o_\b
 =
 \frac{1}{32}
 \o \g^i \g^j \o \, (\g^i \g^j)_{\a\b}
 +
 \frac{1}{96}
 \o \g^i \g^j \g^k \o \, (\g^i \g^j \g^k)_{\a\b}
 \label{Fierz}
\end{equation}
There are two identities, quartic in $\o$, used to show that the
$\o^4$ terms cancel.
\begin{equation}
\begin{split}
 \o \g^i \g^j \o \, \o \g^j \g^i \o
 &=
 \frac{1}{32} \o \g^a \g^b \o \, \o \g^i \g^j \g^a \g^b \g^j \g^i \o
 + \frac{1}{96}
  \o \g^a \g^b \g^c \o \, \o \g^i \g^j \g^a \g^b \g^c \g^j \g^i \o
 \\
 &=
 \frac{25}{32} \o \g^a \g^b \o \, \o \g^a \g^b \o
 + \frac{9}{96} \o \g^a \g^b \g^c \o \, \o \g^a \g^b \g^c \o
 \\
 &=
 \frac{25}{32} \o \g^a \g^b \o \, \o \g^a \g^b \o
 - \frac{9}{32} \o \g^a \g^b \o \, \o \g^a \g^b \o
 \\
 &=
 -\frac{1}{2} \o \g^i \g^j \o \, \o \g^j \g^i \o
 \\
 &=
 0,
\end{split}
\label{identity41}
\end{equation}
where we have used \eqref{Fierz} twice.
Similarly,
\begin{equation}
\begin{split}
 \o \g^i \rs \g^j \o \, \o \g^j \rs \g^i \o
 &=
 \o \rs \g^i \g^j \o \, \o \g^j \g^i \rs \o
 \\
 &=
 \frac{1}{2} \o \g^a \g^b \o \, \o \rs \g^a \g^b \rs \o
 \\
 &=
 2 \o \rs \g^i \o \, \o \g^i \rs \o.
 \label{identity42}
\end{split}
\end{equation}

Calculating the $\o^8$ term requires a number of identities.
First, some more that are quartic in $\o$.
\begin{equation}
 \begin{split}
 \o \g^i \rs \g^j \o \, \o \g^j \rs \g^k \o
 &=
 \frac{5}{32}
 \o \g^{ab} \o \, \o \g^i \rs \g^{ab} \rs \g^k \o
 -
 \frac{3}{96}
 \o \g^{abc} \o \, \o \g^i \rs \g^{abc} \rs \g^k \o
 \\
 &=
 \frac{1}{4}
 \o \g^{ab} \o \, \o \g^i \rs \g^{ab} \rs \g^k \o
 - 3 \o \g^i \rs \o \, \o \rs \g^k \o
 \\
 &=
 r^2 \o \g^{ia} \o \, \o \g^{ak} \o
 - \o \rs \g^a \o \, \o \g^i \rs \g^a \g^k \o
 - 3 \o \g^i \rs \o \, \o \rs \g^k \o
 \\
  &=
  r^2 \o \g^{ia} \o \, \o \g^{ak} \o
  - 5 \o \g^i \rs \o \, \o \rs \g^k \o
  + \o \rs \g^a \o \, \o \g^a \rs \o \, \d^{ik}
  \\ & \qquad
  - r^i \, \o \rs \g^a \o \, \o \g^a \g^k \o
  - \o \g^i \g^a \o \, \o \g^a \rs \o \, r^k
 \end{split}
 \label{identity43}
\end{equation}
Multiplying \eqref{identity43} by $\d^{ik}$ gives
\eqref{identity42}.  Finally,
\begin{equation}
 \begin{split}
 \o \g^i \g^j \o \, \o \g^j \rs \g^k \o
 &=
 \o \g^{ab} \o \, \o \g^i \g^{ab} \rs \g^k \o
 \\
 &=
 \o \g^i \rs \g^j \o \, \o \g^j \g^k \o
 -
 \o \rs \g^a \o \, \o \g^i \g^a \g^k \o.
 \end{split}
\end{equation}

Now some that are octic in $\o$.
\begin{equation}
 \begin{split}
  \o \rs \g^i \o \,
  \o \g^i \rs \g^j \o \,
  \o \g^j \rs \g^k \o \,
  \o \g^k \rs \o
  &=
  \o \rs \g^i \o \,
  \bigl(
   r^2 \o \g^{ia} \o \, \o \g^{ak} \o
   - 5 \o \g^i \rs \o \, \o \rs \g^k \o
   + \o \rs \g^a \o \, \o \g^a \rs \o \, \d^{ik}
   \mskip -20mu
   \\ & \qquad
   - r^i \, \o \rs \g^a \o \, \o \g^a \g^k \o
   - \o \g^i \g^a \o \, \o \g^a \rs \o \, r^k
  \bigr) \,
 \o \g^k \rs \o
 \\
 &=
 r^2
 \o \rs \g^i \o \,
 \o \g^{ia} \o \,
 \o \g^{ak} \o
 \o \g^k \rs \o
 -
 4
 (\o \rs \g^i \o \, \o \g^i \rs \o)^2
 \end{split}
 \label{identity81}
\end{equation}
Similarly,
\begin{equation}
 \begin{split}
  \o \g^i \g^j \o \,
  \o \g^j \rs \g^k \o \,
  \o \g^k \rs \g^l \o \,
  \o \g^l \g^i \o
  &=
  r^2
  \o \g^i \g^j \o \,
  \o \g^{ja} \o \, \o \g^{al} \o
  \o \g^l \g^i \o
  \\ & \qquad
  -
  7
  \o \rs \g^i \o \,
  \o \g^i \g^j \o \,
  \o \g^j \g^k \o \,
  \o \g^k \rs \o,
 \end{split}
 \label{identity82}
\end{equation}
\begin{equation}
 \begin{split}
  \o \g^i \rs \g^j \o \,
  \o \g^j \rs \g^k \o \,
  \o \g^k \rs \g^l \o \,
  \o \g^l \rs \g^i \o
  &=
  r^4
  \o \g^i \g^j \o \,
  \o \g^{ja} \o \, \o \g^{al} \o
  \o \g^l \g^i \o
  \\ & \qquad
  -
  12 r^2
  \o \rs \g^i \o \,
  \o \g^i \g^j \o \,
  \o \g^j \g^k \o \,
  \o \g^k \rs \o
  \\ & \qquad\qquad
  +
  22 (\o \rs \g^i \o \, \o \g^i \rs \o)^2.
 \end{split}
 \label{identity83}
\end{equation}
The identity
\begin{equation}
 \begin{split}
  \o \g^i \g^j \o \,
  \o \g^j \rs \g^k \o \,
  \o \g^k \g^a \g^i \o \,
  \o \rs \g^a \o
  &=
  -\o \g^k \g^j \o \,
  \o \g^j \rs \g^i \o \,
  \o \g^i \g^a \g^k \o \,
  \o \rs \g^a \o
  \\ & \qquad
  -
  \o \rs \g^b \o \,
  \o \g^i \g^b \g^k \o \,
  \o \g^k \g^a \g^i \o \,
  \o \rs \g^a \o
 \\
 &=
 -
 \frac{1}{2}
 \o \rs \g^b \o \,
 \o \g^i \g^b \g^k \o \,
 \o \g^k \g^a \g^i \o \,
 \o \rs \g^a \o
 \\
 &=
 \o \rs \g^b \o \,
 \o \g^b \g^k \o \,
 \o \g^k \g^a \o \,
 \o \g^a \rs \o
 \end{split}
\end{equation}
is needed for
\begin{equation}
 \begin{split}
  \o \g^i \g^j \o \,
  \o \g^j \rs \g^k \o \,
  \o \g^k \g^l \o \,
  \o \g^l \rs \g^i \o
  &=
  r^2
  \o \g^i \g^j \o \,
  \o \g^{ja} \o \, \o \g^{al} \o
  \o \g^l \g^i \o
  \\ & \qquad
  -
  8
  \o \rs \g^i \o \,
  \o \g^i \g^j \o \,
  \o \g^j \g^k \o \,
  \o \g^k \rs \o.
 \end{split}
 \label{identity84}
\end{equation}

\begin{flushleft}
 \bibliographystyle{hieeetr}
 \bibliography{Matrix}
\end{flushleft}

\end{fmffile}

\end{document}